\documentclass[aps,prd,twocolumn,showpacs,floats,floatfix,letterpaper,nofootinbib,superscriptaddress,]{revtex4}

\usepackage{amssymb,amsmath,latexsym,mathrsfs}
\usepackage{graphicx}
\usepackage{epsfig}
\usepackage{multirow}
\usepackage{array}

\begin{document}


\title{Constraints on massive sterile plus active neutrino species in non minimal cosmologies}

\author{Elena Giusarma}
\affiliation{IFIC, Universidad de Valencia-CSIC, 46071, Valencia, Spain}
\author{Maria Archidiacono}
\affiliation{Physics Department and INFN, Universita' di Roma 
	``La Sapienza'', Ple.\ Aldo Moro 2, 00185, Rome, Italy}
\author{Roland de Putter$^1$}
\affiliation{ICC, University of Barcelona (IEEC-UB), Marti i Franques 1, Barcelona 08028, Spain}
\author{Alessandro Melchiorri$^2$}
\author{Olga Mena$^1$}

\begin{abstract}
Cosmological measurements are affected by the energy density of both active and sterile massive neutrinos. We extend here a recent analysis of current cosmological data to non minimal cosmologies. Several possible scenarios are examined: a constant $w \neq -1$ dark energy equation of state, a non flat universe, a time varying dark energy component and coupled dark matter dark energy universes or modified gravity scenarios. When considering cosmological data only, (3+2) massive neutrino models with $\sim 0.5$~eV sterile species are allowed at $95\%$ CL. This scenario has been shown to reconcile reactor, LSND and MiniBooNE positive signals with null results from other searches. Big Bang Nucleosynthesis bounds could compromise the viability of (3+2) models if the two sterile species are fully thermalized states at decoupling.
\end{abstract}

\pacs{98.80.-k 95.85.Sz,  98.70.Vc, 98.80.Cq}

\maketitle

\section{Introduction}

Solar, atmospheric, reactor, and accelerator neutrinos have provided compelling evidence for the existence of neutrino oscillations, implying non-zero neutrino masses (see Ref.~\cite{GonzalezGarcia:2007ib} and references therein). The present data require the number of massive neutrinos to be equal or larger than two, since there are at least two mass squared differences ($\Delta m^2_{\rm atmos}$ and $\Delta m^2_{\rm solar}$) driving the atmospheric and solar neutrino oscillations respectively. Unfortunately, oscillation experiments only provide bounds on the neutrino mass squared differences, i.e. they are not sensitive to the overall neutrino mass scale.

Cosmology provides one of the means to tackle the absolute scale of neutrino
masses. Neutrinos can leave key signatures in several cosmological data sets. 
The amount of primordial relativistic neutrinos changes the epoch of the matter-
radiation equality, leaving an imprint on both Cosmic Microwave Background (CMB) anisotropies (through the so-called Integrated Sachs-Wolfe effect) and on 
structure formation, while non relativistic neutrinos in the recent Universe 
suppress the growth of matter density fluctuations and galaxy clustering, see
Ref.~\cite{Lesgourgues:2006nd}. Cosmology can therefore weigh neutrinos, providing an upper bound on the sum of the three active neutrino masses, $\sum m_\nu \sim 0.58$~eV at $95\%$ CL~\cite{Komatsu:2010fb}. The former bound is found when CMB measurements from the Wilkinson Microwave Anisotropy Probe (WMAP) are combined with measurements of the distribution of galaxies (SDSS-II-BAO) and of the Hubble constant $H_0$ (HST)~\footnote{For other recent analyses, see also Refs.~\cite{Reid:2009nq,Hamann:2010pw}.} in the assumption of a flat universe with a cosmological constant, i.e. a $\Lambda$CDM cosmology.

There is no fundamental symmetry in nature forcing  a definite number of right-handed (sterile) neutrino species, as those are allowed in the Standard Model fermion content. Indeed, cosmological probes have been extensively used to set bounds on the relativistic energy density of the universe in terms of the effective number of neutrinos $N_\nu^{\rm eff}$ (see, for instance, 
Refs.~\cite{Mangano:2006ur,Hamann:2007pi,Reid:2009nq,Hamann:2010pw,Mangano:2010ei,Komatsu:2010fb,darkr}. If the effective number of neutrinos $N_\nu^{\rm eff}$ is larger than the Standard Model prediction of $N_\nu^{\rm eff}=3.046$ at the Big Bang Nucleosynthesis (BBN) era, the relativistic degrees of freedom, and, consequently, the Hubble expansion rate will also be larger causing weak interactions to become uneffective earlier. This will lead to a larger neutron-to-proton ratio and will change the standard BBN predictions for light element abundances. Combining Deuterium and $^4$He data, the authors of Ref.~\cite{Mangano:2006ur} found $N_\nu^{\rm eff}=3.1^{+1.4}_{-1.2}$ at $95\%$ CL. 

Models with one additional $\sim 1$~eV massive sterile neutrino, i.e. the so called (3+1) models, were introduced to explain LSND short baseline (SBL) antineutrino data~\cite{Aguilar:2001ty} by means of neutrino oscillations. A much better fit to SBL appearance data and, to a lesser extent, to disappearance data, is provided by models with two sterile neutrinos (3+2)~\cite{Sorel:2003hf,Karagiorgi:2006jf} which can also explain both the MiniBooNE neutrino~\cite{AguilarArevalo:2007it} and antineutrino data~\cite{AguilarArevalo:2009xn} if CP violation is allowed~\cite{Karagiorgi:2009nb}. More recently, a combined analysis including the new reactor antineutrino fluxes~\cite{Mueller:2011nm,Mention:2011rk} has shown that (3+2) models provide a very good fit to short baseline data~\cite{Kopp:2011qd}. While these models with extra sterile species show some tension with BBN bounds on $N_\nu^{\rm eff}$, the extra sterile neutrinos do not necessarily have to feature thermal abundances at decoupling, see Refs.~\cite{Melchiorri:2008gq,Acero:2008rh}, where the usual full thermalization scenario for the sterile neutrino species was not assumed. Up to date cosmological constraints on massive sterile and active neutrino species have been presented in Refs.~\cite{rt,us} in the context of a $\Lambda$CDM universe. It is well known that bounds on active neutrino species are relaxed if the dark energy equation of state is different from $-1$~\cite{Hannestad:2005gj,Hamann:2011ge,Calabrese:2011hg} and/or interactions between the dark matter and dark energy sectors are switched on~\cite{LaVacca:2008mh,Gavela:2009cy}. In the same line, the authors of Ref.~\cite{Kristiansen:2011mp} have found that in models with non zero curvature and two extra sterile neutrinos the cosmological constant scenario is ruled out at $95\%$ CL.

Here we extend the minimal cosmological scenario considered in our previous study~\cite{us} and compute the bounds on the masses of the active and the sterile neutrino states as well as on the number of sterile states in the presence of a constant equation of state $w\neq1$, a time varying dark energy fluid, a non vanishing curvature component and interactions among the dark sectors. The paper is organized as follows. Section \ref{sec:i} describes the details of the analysis carried out here, including the cosmological parameters and datasets. The four different cosmological scenarios explored here are analyzed and the most important degeneracies among the neutrino parameters are carefully explored. Section \ref{sec:seciii} summarizes our main results and conclusions.

\section{Cosmological constraints}
\label{sec:i}
Here we present the constraints from current data on the active neutrino masses and on the sterile neutrino thermal abundance and masses in different cosmological scenarios. We have modified the Boltzmann CAMB code~\cite{camb} incorporating the extra massive sterile neutrino parameters and extracted cosmological parameters from current data using a Monte Carlo Markov Chain (MCMC) analysis based on the publicly available MCMC package \texttt{cosmomc}~\cite{Lewis:2002ah}. We consider here four possible scenarios: the $w$CDM model in which we include the possibility of a dark energy equation of state parameter $w$ different from -1; the $w(a)$CDM model in which we assume an equation of state evolving with redshift, the $\Omega_k$CDM model where we allow the spatial curvature of the universe to vary, and the model in which an interaction among the dark matter and dark energy sectors is switched on, the $\xi$CDM model. These scenarios are an extension of the minimal cosmological model plus three ($N_{\nu_s}$) active (sterile) massive neutrino species.  We consider subsets of the following parameters:
\begin{widetext}
 \label{parameter}
  $\{\omega_b,\omega_c, \Theta_s, \tau, n_s, \log[10^{10}A_{s}], m_\nu , m_{\nu_s}, N_{\nu_s}, w (w_0), w_a, \Omega_k,\xi\}$~,
\end{widetext}
where $\omega_b\equiv\Omega_bh^{2}$ and $\omega_c\equiv\Omega_ch^{2}$ are the physical baryon and cold dark matter densities, $\Theta_{s}$ is the ratio between the sound horizon and the angular diameter distance at decoupling, $\tau$ is the optical depth, $n_s$ is the scalar spectral index, $A_{s}$ is the amplitude of the primordial spectrum, $m_\nu$ is the active neutrino mass, $m_{\nu_s}$ is the sterile neutrino mass and $N_{\nu_s}$ is the number of thermalized sterile neutrino species\footnote{We assume that both active and sterile neutrinos have a degenerate mass spectra.}, $w$ is the dark energy equation of the state parameter, $\Omega_k$ is the curvature parameter, $\xi$ is the dimensionless parameter which encodes the dark matter dark energy interaction, and $w_0$, $w_a$ are parameters related to the dark energy equation of state. Table \ref{tab:priors} specifies the priors considered on the different cosmological parameters. In all cases in which $w\neq -1$ we have considered the effect of dark energy perturbations, fixing the dark energy speed of sound $c^2_s=1$. 

Our basic data set is the seven--year WMAP data \cite{Komatsu:2010fb,wmap7}  (temperature and polarization) with the routine for computing the likelihood supplied by the WMAP team. We consider two cases: we first analyze the WMAP data together with the luminous red galaxy clustering results from SDSS-II (Sloan Digital Sky Survey)~\cite{beth} and with a prior on the Hubble constant from HST (Hubble Space Telescope)~\cite{Riess:2009pu}, referring to it as the  ``run1'' case. We then include with these data sets Supernova Ia Union Compilation 2 data~\cite{sn}, and we will refer to this case as ``run2''. In addition, we also add to the previous data sets the BBN measurements of the $^4$He abundance, considering separately helium fractions of $Y^1_p=0.2561\pm 0.0108$ (see Ref.~\cite{aver}) and of $Y^2_p=0.2565\pm 0.0010$ (stat.) $\pm 0.0050$ (syst.) from Ref.~\cite{it}. Finally, we also consider the Deuterium abundance measurements $\log (D/H) = -4.56 \pm  0.04$ from Ref.~\cite{pettini}.

Since the CMB data we are considering are not significantly
constraining the amount of primordial Helium abundance, 
 we  fix it  to the value $Y_p=0.24$, consistent with current observations.
Then we use the MCMC chains from each different run and perform importance sampling
obtaining the predicted values for $Y_p$ and $\log (D/H)$ with an interpolation routine using a grid of the public available PArthENoPE BBN code (see \cite{iocco}) for each point ($\omega_b$, $N_\nu^{\rm eff}=3+N_{\nu_s}$) of a given cosmological model, as in \cite{hamann2}.

In the following, we will present the cosmological constraints on the masses of the active and the sterile neutrino states as well as on the number of sterile states for different cosmological scenarios, namely,  a universe with a constant equation of state $w \neq -1$,  a universe with a time varying dark energy fluid, a universe with a non vanishing curvature component and a universe with interacting dark matter-dark energy sectors.

\begin{table}[h!]
\begin{center}
\begin{tabular}{c|c}
\hline\hline
 Parameter & Prior\\
\hline
$\Omega_{b}h^2$ & $0.005 \to 0.1$\\
$\Omega_{c}h^2$ & $0.01 \to 0.99$\\
$\Theta_s$ & $0.5 \to 10$\\
$\tau$ & $0.01 \to 0.8$\\
$n_{s}$ & $0.5 \to 1.5$\\
$\ln{(10^{10} A_{s})}$ & $2.7 \to 4$\\
$m_{\nu_s}$\ [eV] &  $0 \to 3$\\
$m_{\nu}$\ [eV] &  $0 \to 3$\\
$N_{\nu_s}$ &  $0 \to 6$\\
$w (w_0)$ & $-2 \to 0$\\
$w_a$ &  $-1 \to  1$\\
$\Omega_{k}$ &  $-0.02 \to  0.03$\\
$\xi$ &  $-2 \to  0.$\\
\hline\hline
\end{tabular}
\caption{Flat priors for the cosmological parameters considered here.}
\label{tab:priors}
\end{center}
\end{table}

\begin{table*}[ht]
\caption{1D marginalized 95\% CL bounds on $N_{\nu_s}$, $m_{\nu_s}$ and $m_\nu$  using the two combinations of data sets described in the text (r1 refers to ``run 1'' and r2 refers to ``run 2'', respectively) for the $w$CDM cosmology. We also show the constraints after combining the results of ``run 2'' with those coming from different measurements of BBN light element abundances.}
\label{tab:bfnu}
\begin{center}\tabcolsep=2mm
\begin{tabular}{c c c c c c c}
\hline\hline
Parameter &  95\% CL (r1) &  95\% CL (r2) &  $Y^1_p$     &     $Y^2_p$     &     $Y^1_p+D$     &     $Y^2_p+D$  \\
\hline
$N_{\nu_s}$ & $<4.4$ & $<3.9$ & $<2.3$   &   $<1.6$   &   $<1.6$ & $<1.3$\\
$m_{\nu}$\ [eV] & $<0.34$ & $<0.31$ & $<0.25$   &   $<0.23$   &   $<0.23$ &  $<0.23$\\
$m_{\nu_s}$\ [eV] & $<0.51$ & $<0.57$ & $<0.68$   &   $<0.71$   &   $<0.75$ & $<0.75$\\
\hline\hline
\end{tabular}
\end{center}
\end{table*}

\subsection{$w$CDM cosmology}
\begin{figure}[h]
\begin{tabular}{c}
\includegraphics[width=7.5cm]{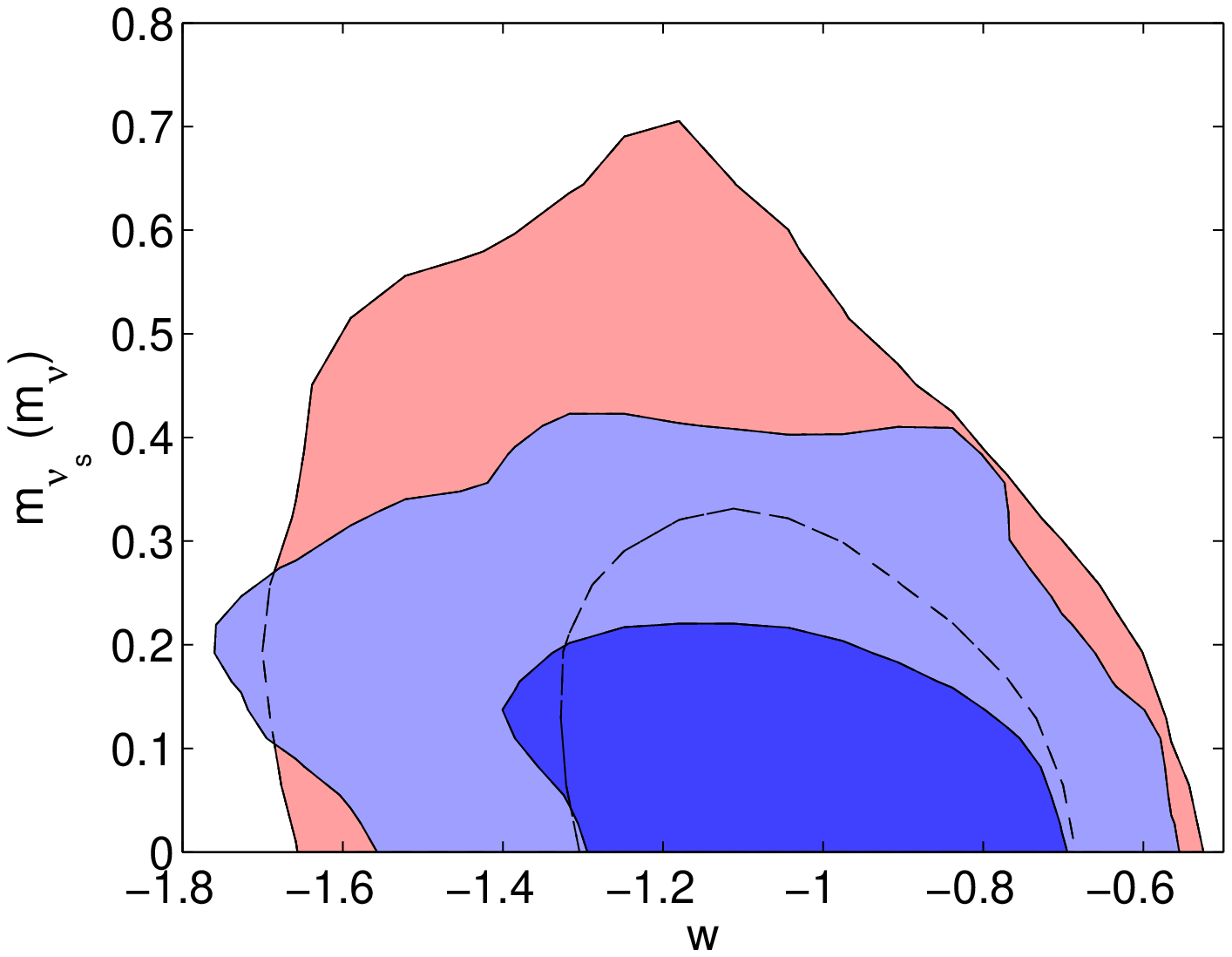}\\
\includegraphics[width=7.5cm]{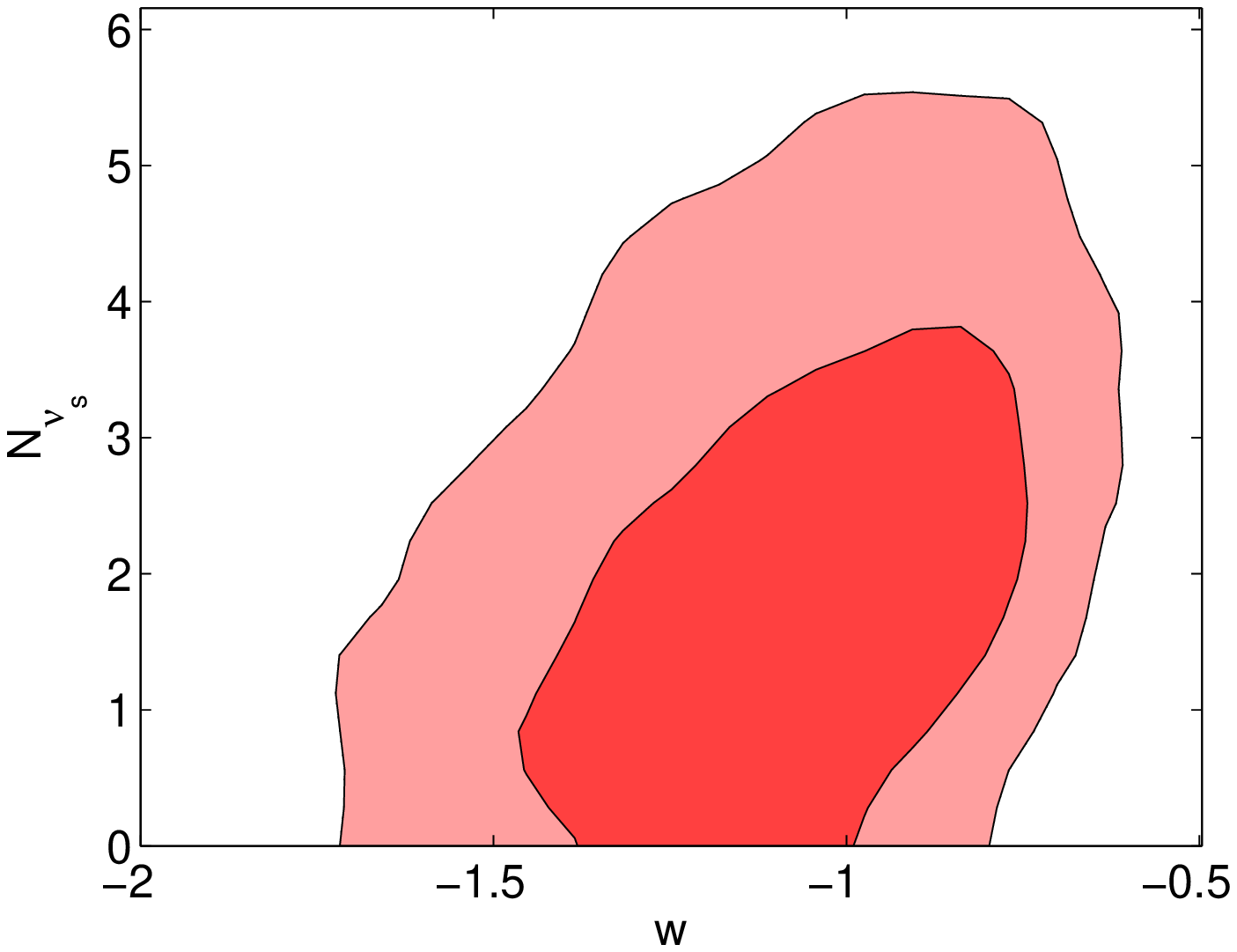}\\
\end{tabular}
\caption{The upper panel shows the 68\%  and 95\% CL bounds from ``run 1'' in the plane $w-m_{\nu}$ (in blue) and $w$-$m_{{\nu}_s}$ (in red), respectively. The masses of the sterile and active neutrinos are both in eV.  The lower panel depicts the 68\%  and 95\% CL contours in the plane $w-N_{{\nu}_s}$ for a constant dark energy equation of state.} 
\label{fig:w_mnus}
\end{figure}

We first consider a cosmological model including standard cold dark matter and a dark energy  fluid characterized by a constant equation of state $w$ .
We consider the following set of parameters:
\begin{equation}
\{\omega_b,\omega_c, \Theta_s, \tau, n_s, \log[10^{10}A_{s}], m_\nu , m_{\nu_s}, N_{\nu_s}, w\}~.
\end{equation}
Table \ref{tab:bfnu} shows the one dimensional (1D) marginalized 95\% CL bounds on $N_{\nu_s}$, $m_{\nu_s}$ and $m_\nu$  using the two combinations of data sets described above. Note that the addition of SNIa data affects only the number of massive sterile neutrino species, and not their masses. 
The bounds obtained in a $\Lambda$CDM scenario, see Tab.~\ref{tab:LCDM} in Appendix~\ref{sec:lcdm}~\cite{us} are slightly relaxed when the dark energy equation of state is allowed to vary. There is a strong and very well known degeneracy in the  $m_\nu-w$ plane (and therefore also in the  $m_{\nu_s}-w$ plane) as first noticed in Ref.~\cite{Hannestad:2005gj}. Cosmological neutrino mass bounds become weaker if the dark energy equation of state is taken as a free parameter. If $w$ is allowed to vary, the cold dark matter mass energy density $\Omega_{c}$ can take very high values, as required when $m_\nu$ (or $m_{\nu_s}$) is increased in order to have the same matter power spectrum. We can observe this degeneracy in the upper panel of Fig.~\ref{fig:w_mnus}. There exists also a degeneracy between the number of sterile neutrino species and the dark energy equation of state, see Fig.~\ref{fig:w_mnus} (lower panel). Sub-eV massive sterile neutrino species may be quasi relativistic states at decoupling. One of the main effects of $N_{\nu_s}$ comes from the change of the epoch of the radiation matter equality, and consequently, from the shift of the CMB acoustic peaks, see Ref.~\cite{Hou:2011ec} for a detailed study. The position of acoustic peaks is given by the so-called acoustic scale $\theta_A$, which reads
\begin{equation}
\theta_A=\frac{r_s(z_{rec})}{r_\theta(z_{rec})}~,
\end{equation}
where $r_\theta (z_{rec})$ and $r_s(z_{rec})$ are the comoving angular diameter distance to the last scattering surface and the sound horizon at the recombination epoch $z_{rec}$, respectively. Although 
$r_\theta (z_{rec})$ almost remains the same for different values of $N_{\nu_s}$, $r_s(z_{rec})$ becomes smaller when $N_{\nu_s}$ is increased. Thus the positions of acoustic peaks are shifted to higher multipoles (smaller angular scales) by increasing the value of $N_{\nu_s}$~\cite{arXiv:0803.0889}. A dark energy component with $w> -1$ will decrease the comoving angular diameter distance to the last scattering surface $r_\theta (z_{rec})$, shifting the positions of the CMB acoustic peaks to larger angular scales, i.e. to lower multipoles $\ell$, compensating, therefore, the effect induced by an increase of $N_{\nu_s}$. 

We have also computed the constraints after combining the results of ``run 2'' with those coming from different measurements of BBN light element abundances, see Tab.~\ref{tab:bfnu}.  The bounds on $N_{\nu_s}$ obtained in a $w$CDM cosmology when adding BBN constraints are very similar to those obtained in a $\Lambda$CDM scenario~\cite{us}. The limits on the active and neutrino masses are mildly relaxed.

\subsection{$w(a)$CDM cosmologies}
\begin{figure}[h]
\begin{tabular}{c}
\includegraphics[width=7.5cm]{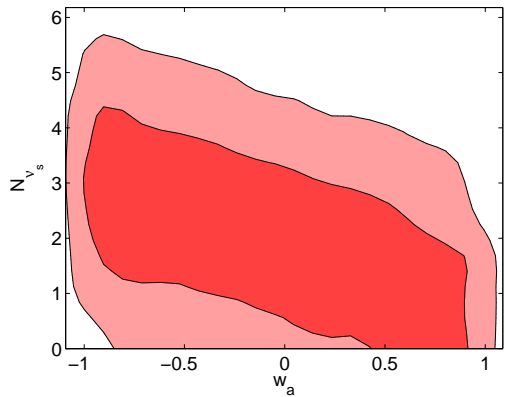}\\
\end{tabular}
\caption{68\%  and 95\% CL contours arising from "run 1" in the plane $w-N_{{\nu}_s}$ for the $w(a)$CDM cosmology.} 
\label{fig:wapart}
\end{figure}

We also consider a time varying equation of state with a parameterization that has been extensively explored in the literature~\cite{Chevallier:2000qy,Linder:2002et,Albrecht:2006um,Linder:2006sv}: 
\begin{equation}
w(a)=w_0+w_a(1-a)~.
\end{equation}
We consider the following set of parameters:
\begin{equation}
\{\omega_b,\omega_c, \Theta_s, \tau, n_s, \log[10^{10}A_{s}], m_\nu , m_{\nu_s}, N_{\nu_s}, w_0, w_a\}~.
\end{equation}
Table \ref{tab:waCDM} shows the 1D marginalized 95\% CL bounds on $N_{\nu_s}$, $m_{\nu_s}$ and $m_\nu$  using the two combinations of data sets used along this manuscript. The addition of SNIa data does not improve at all the constraints from "run 1". There exist large degeneracies in the $m_\nu-w_0$, $m_{\nu_s}-w_0$ and  $N_{\nu_s}-w_0$ planes. These degeneracies are  identical to the $w$CDM cosmology ones, and therefore we will not illustrate them here to avoid redundancy. We show instead the mild degeneracy in the $N_{\nu_s}-w_a$ plane, see  Fig.~\ref{fig:wapart}. In this case, an increase of $N_{\nu_s}$ will be compensated by  a decrease of $w_a$. The BBN bounds combined with the "run 2" constraints on the neutrino parameters are also depicted in Tab. ~\ref{tab:waCDM}.

\begin{table*}[ht]
\caption{As Tab.~\ref{tab:bfnu}, but for the $w(a)$CDM cosmology.}
\label{tab:waCDM}
\begin{center}\tabcolsep=2mm
\begin{tabular}{c c c c c c c}
\hline\hline
Parameter &  95\% CL(r1) &  95\% CL(r2) &  $Y^1_p$     &     $Y^2_p$     &     $Y^1_p+D$     &     $Y^2_p+D$  \\
\hline
$N_{\nu_s}$ & $<4.5$ & $<4.4$ & $<2.6$   &   $<1.7$   &   $<1.7$ & $<1.4$\\
$m_{\nu}$\ [eV] & $<0.33$ & $<0.31$ & $<0.26$   &   $<0.23$   &   $<0.22$ &  $<0.22$\\
$m_{\nu_s}$\ [eV] & $<0.49$ & $<0.48$ & $<0.58$   &   $<0.66$   &   $<0.73$ & $<0.72$\\\hline\hline
\end{tabular}
\end{center}
\end{table*}

\subsection{Non flat $\Omega_k$CDM cosmologies}
We also explore here the constraints in non flat cosmologies. 
We consider the following set of parameters:
\begin{equation}
\{\omega_b,\omega_c, \Theta_s, \tau, n_s, \log[10^{10}A_{s}], m_\nu , m_{\nu_s}, N_{\nu_s}, \Omega_k\}~.
\end{equation}
Current CMB measurements combined with SNIa and BAO data, give the constraint $\Omega_k=-0.0057^{+0.0067}_{-0.0068}$~\cite{Komatsu:2010fb}. Table \ref{tab:omkCDM} shows the 1D marginalized 95\% CL bounds on $N_{\nu_s}$, $m_{\nu_s}$ and $m_\nu$  using the two combinations of data sets considered here. The bounds are very similar to those obtained in the other two cosmological models explored above. 

\begin{figure}[h]
\begin{tabular}{c}
\includegraphics[width=7.5cm]{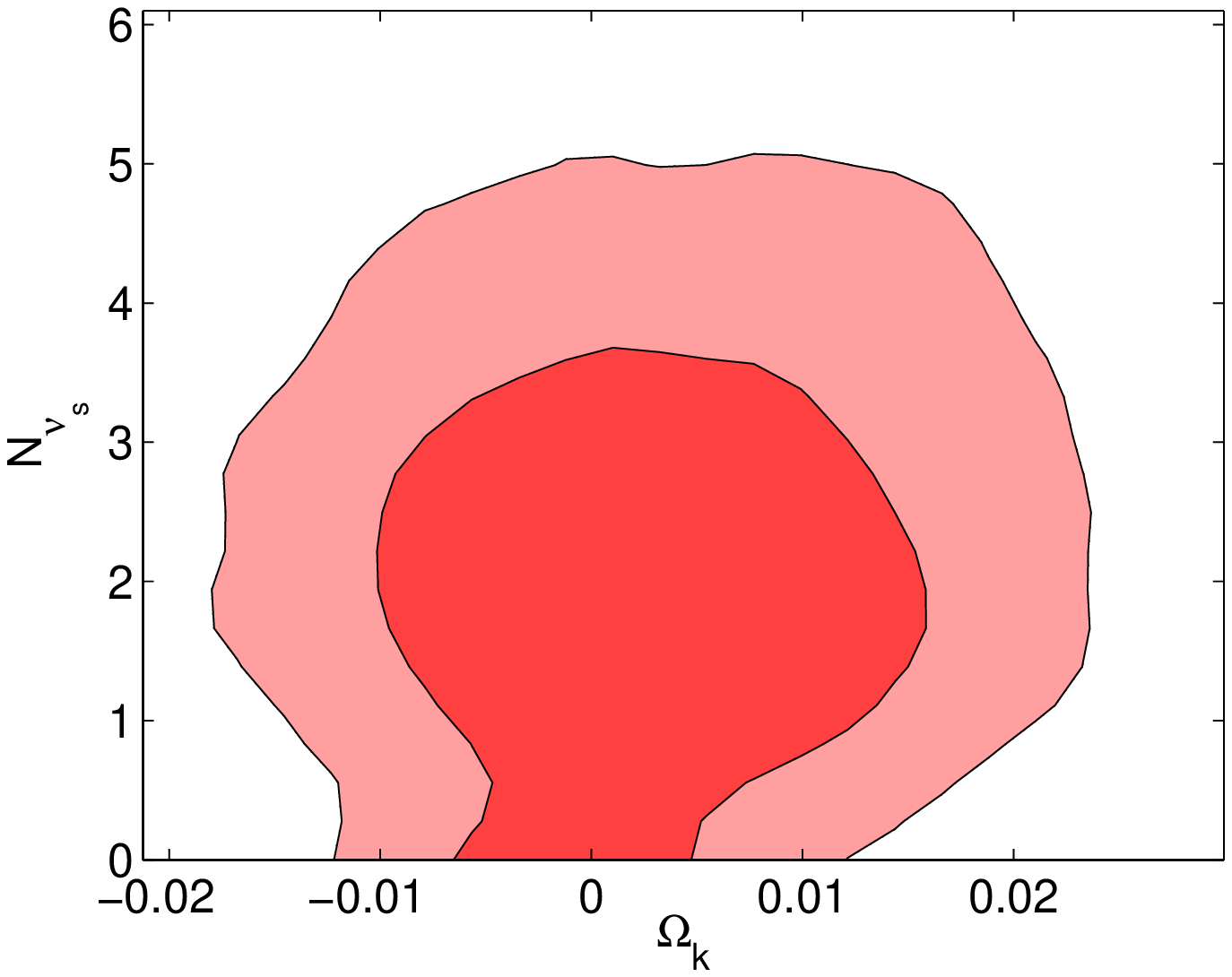}\\
\includegraphics[width=7.5cm]{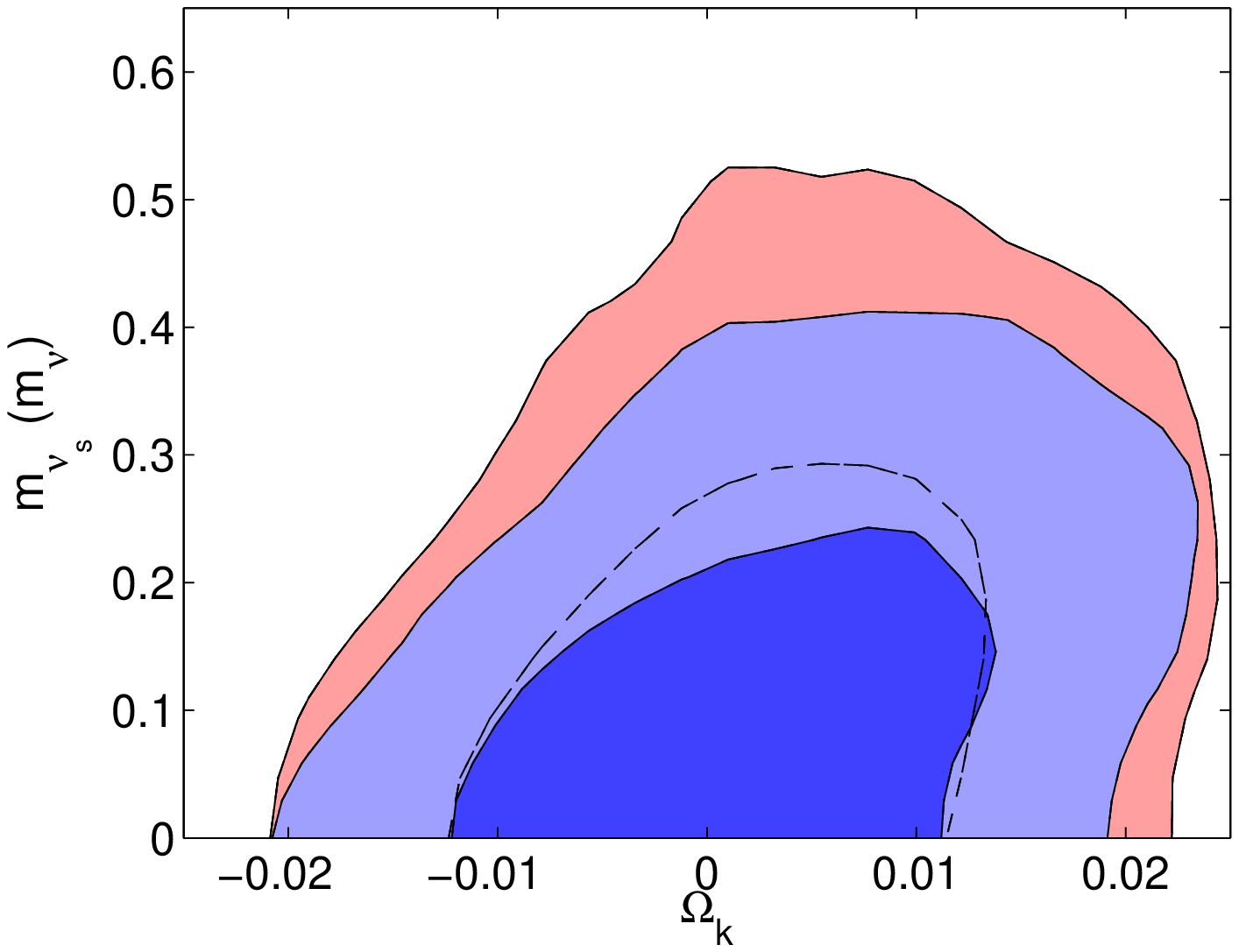}\\
\end{tabular}
\caption{The upper panel shows the 68\%  and 95\% CL bounds from ``run 1'' in the $\Omega_k-N_{{\nu}_s}$ plane. The lower plane shows the analogous in the $\Omega_k-m_{\nu}$ (in blue) and $\Omega_k-m_{{\nu}_s}$ (in red) planes, respectively. The masses of the sterile and active neutrinos are both in eV.} 
\label{fig:omk_mnus}
\end{figure}

Figure \ref{fig:omk_mnus} shows the existing degeneracies between the curvature energy density and the number of massive sterile neutrino species $N_{\nu_s}$ and between the curvature and the masses of the active and sterile neutrino species. In an open universe with $\Omega_k>0$ 
 the comoving angular diameter distance to the last scattering surface $r_\theta (z_{rec})$ will be higher, shifting the positions of the CMB acoustic peaks to smaller angular scales, i.e. to larger multipoles $\ell$, which can be compensated with a larger dark matter energy density which will allow for higher neutrino masses. At the same time a higher dark matter energy density will decrease the height of the acoustic peaks, 
features which can be compensated by a larger number of sterile neutrino species.

Table \ref{tab:omkCDM} also show the constraints arising from BBN measurements 
combined with "run 2" bounds.

\begin{table*}[ht]
\caption{As Tab.~\ref{tab:bfnu}, but for the $\Omega_k$CDM cosmology.}
\label{tab:omkCDM}
\begin{center}\tabcolsep=2mm
\begin{tabular}{c c c c c c c}
\hline\hline
Parameter &  95\% CL(r1) &  95\% CL(r2) &  $Y^1_p$     &     $Y^2_p$     &     $Y^1_p+D$     &     $Y^2_p+D$  \\
\hline
$N_{\nu_s}$ & $<4.3$ & $<3.8$ & $<2.4$   &   $<1.7$   &   $<1.7$ & $<1.4$\\
$m_{\nu}$\ [eV] & $<0.34$ & $<0.31$ & $<0.24$   &   $<0.22$   &   $<0.22$ &  $<0.21$\\
$m_{\nu_s}$\ [eV] & $<0.45$ & $<0.52$ & $<0.70$   &   $<0.72$   &   $<0.77$ & $<0.75$\\
\hline\hline
\end{tabular}
\end{center}
\end{table*}

\subsection{Coupled $\xi$CDM cosmologies and modified gravity scenarios}
Interactions within the dark sectors, i.e. between cold dark matter and dark energy, are still allowed by 
observations (see Ref.~\cite{Honorez:2010rr} and references therein).  We parametrize the dark matter-dark  energy interactions at the level of the stress-energy tensor conservation equations, introducing 
an energy momentum exchange of the following form~\cite{Gavela:2010tm}:
\begin{eqnarray}
\nabla_\mu T^\mu_{({c})\nu} &=&Q_\nu~;\nonumber \\
\nabla_\mu T^\mu_{({de})\nu} &=&-Q_\nu~,
\label{eq:conservDMDE}
\end{eqnarray}
with
\begin{equation}
    Q_\nu= \xi {\mathcal H} \rho_{de} u_{\nu}^{c}/a \qquad\mbox{or}\qquad  
    Q_\nu= \xi {\mathcal H} \rho_{de} u_{\nu}^{de}/a~,
\label{eq:ourm}
\end{equation}
where $ u_{\nu}^{c (de)}$ is the cold dark matter (dark energy) four velocity and $\xi$ is a dimensionless coupling, considered negative in order to avoid early time non adiabatic instabilities~\cite{Gavela:2009cy}. In general, coupled models with $Q_\nu$ proportional to $u_{\nu}^{de}$ are effectively modified gravity models. 
We consider the following set of parameters:
\begin{equation}
\{\omega_b,\omega_c, \Theta_s, \tau, n_s, \log[10^{10}A_{s}], m_\nu , m_{\nu_s}, N_{\nu_s}, w, \xi \}~.
\end{equation}

\begin{figure}[h]
\begin{tabular}{c}
\includegraphics[width=7.5cm]{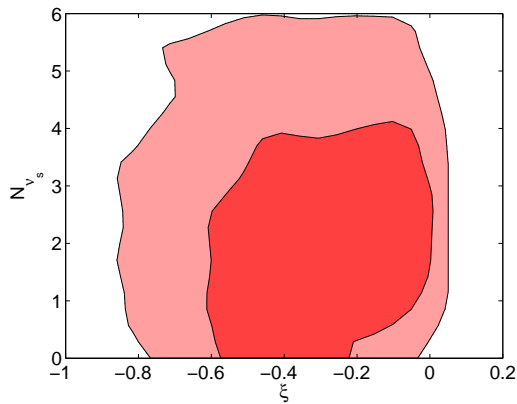}
\end{tabular}
\caption{68\%  and 95\% CL bounds arising from the "run 1" analysis in the plane $\xi-N_{\nu_s}$ for a universe with an interacting dark matter-dark energy fluid.} 
\label{fig:xi}
\end{figure}

Interactions between the dark matter and dark energy sectors can relax the bounds on the active neutrino masses~\cite{LaVacca:2008mh,Gavela:2009cy} since, for negative couplings, the power spectrum increases due the the higher matter energy density  when $\xi < 0$. Such a power enhancement effect (induced by the presence of a coupling) can be compensated by adding massive neutrinos in the game, which will suppress the matter power spectrum. Therefore, there exists a well-known $m_\nu-\xi$ degeneracy. Here we illustrate an additional degeneracy, the one existing in the $N_{\nu_s}-\xi$ plane. Figure \ref{fig:xi} shows the existing degeneracy among the coupling $\xi$ and the number of sterile massive neutrino species for a coupling term proportional to $u_{\nu}^{c}$ (identical results are obtained for the case proportional to $u_{\nu}^{de}$). As first noticed in Ref.~\cite{Gavela:2009cy} a huge degeneracy is present between the coupling $\xi$ and the mass energy density of cold 
 dark matter $\omega_c$, with the former two quantities having positive correlations. In a universe with a negative dark coupling $\xi$, the matter content in the past is higher than in a standard $\Lambda$CDM scenario, since the cold dark matter and dark energy densities read
\begin{eqnarray}
\Omega_c &=&\Omega^{0}_c a^{-3} +\Omega^{0}_{de} \frac{\xi}{(w +\frac{\xi}{3})} \left(1-a^{-3(w+\frac{\xi}{3})}\right) a^{-3}~;\nonumber \\
\Omega_{de} &=&\Omega^{0}_{de} a^{-3(1+w+\frac{\xi}{3})}~, \\
\end{eqnarray}
with $\Omega^{0}_{c,de}$ the current cold dark matter (dark energy) mass energy densities. Therefore, the amount of intrinsic dark matter needed - that is, not including the contribution of dark energy through the coupling term - should decrease as the dark coupling becomes more and more negative. Therefore, the number of effective neutrino species should also decrease as the coupling gets more negative to leave unchanged both the matter-radiation equality epoch and the first CMB peak heigh. Table \ref{tab:xi} shows the analogous to previous sections but for the coupled case.
Notice that the bounds on the number of massive sterile neutrinos arising from "run1" and "run 2" analyses are milder than those obtained in the previous cosmologies, due to the degeneracy between the coupling $\xi$ and $N_{\nu_s}$.
\begin{table*}[ht]
\caption{As Tab.~\ref{tab:bfnu}, but for the $\xi$CDM cosmology.}
\label{tab:xi}
\begin{center}\tabcolsep=2mm
\begin{tabular}{c c c c c c c}
\hline\hline
Parameter &  95\% CL(r1) &  95\% CL(r2) &  $Y^1_p$     &     $Y^2_p$     &     $Y^1_p+D$     &     $Y^2_p+D$  \\
\hline
$N_{\nu_s}$ & $<5.2$ & $<4.7$ & $<2.4$   &   $<1.7$   &   $<1.8$ & $<1.4$\\
$m_{\nu}$\ [eV] & $<0.34$ & $<0.33$ & $<0.21$   &   $<0.21$   &   $<0.20$ &  $<0.20$\\
$m_{\nu_s}$\ [eV] & $<0.35$ & $<0.38$ & $<0.47$   &   $<0.51$   &   $<0.49$ & $<0.49$\\
\hline\hline
\end{tabular}
\end{center}
\end{table*}

\section{Discussion and Conclusions}
\label{sec:seciii}
\begin{figure}
\begin{tabular}{c}
\includegraphics[width=7.5cm]{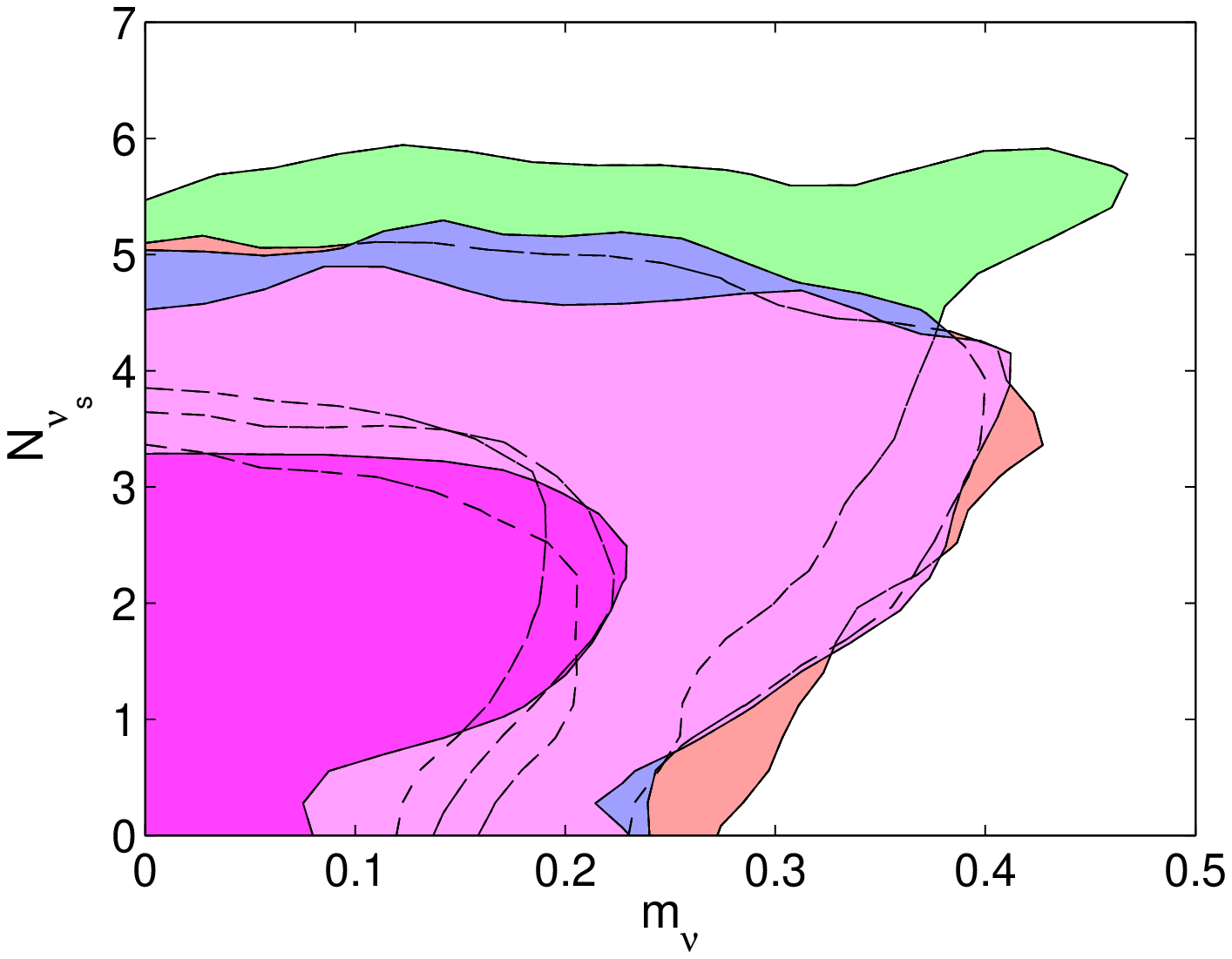}\\
\includegraphics[width=7.5cm]{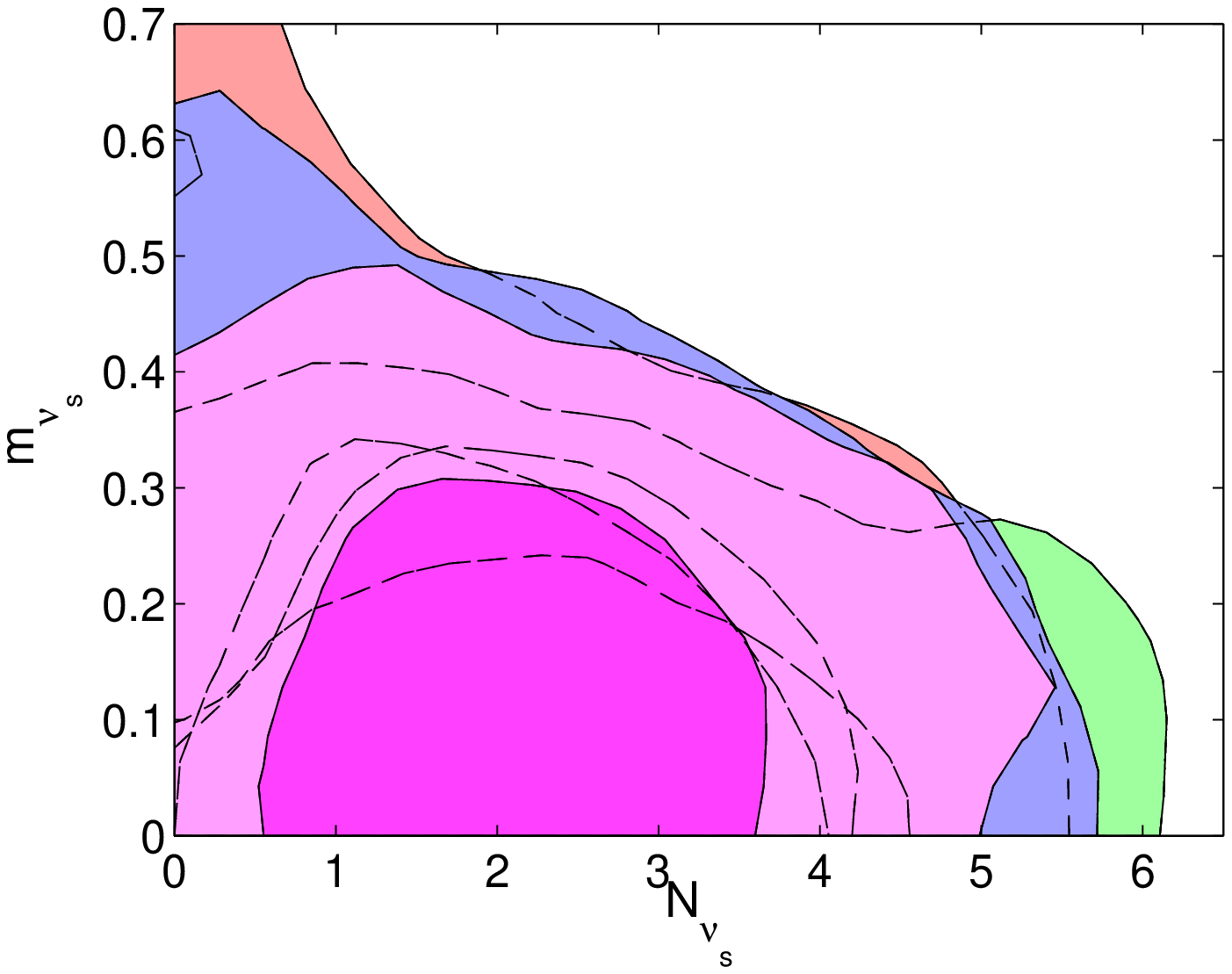}\\
\end{tabular}
\caption{The upper panel shows the 68\%  and 95\% CL bounds from ``run 1'' in the $m_\nu-N_{{\nu}_s}$ plane. The lower plane shows the analogous in the $N_{{\nu}_s}-m_{{\nu}_s}$ plane. The masses of the sterile and active neutrinos are both in eV. Red, blue, magenta and green contours denote $w$CDM, $w(a)$CDM, $\Omega_k$CDM and $\xi$CDM cosmologies, respectively.} 
\label{fig:final}
\end{figure}

Neutrino oscillation experiments indicate that neutrinos have non zero masses and open the possibility for a number of extra sterile neutrino species. LSND and MiniBooNE antineutrino data require these extra sterile species to be massive. Much effort has been devoted in the literature to constrain the so called (3+1) (three active plus one sterile) and (3+2) (three active plus two sterile). Recently, global fit analyses incorporating new reactor antineutrino fluxes have shown that (3+2) models with $0.5-1$~eV sterile species provide a very good fit to short baseline data. Cosmology can set bounds on both the active and sterile neutrino masses as well as on the number of sterile neutrino species. We have explored here the current constraints on these parameters in natural extensions of the minimal $\Lambda$CDM cosmology. Namely, we have explored the neutrino constraints in scenarios without a cosmological constant as the dark energy fluid, with a non vanishing curvature, or with coupled dark matter-dark energy fluids. 
Figure \ref{fig:final} summarizes our results in the $m_\nu-N_{{\nu}_s}$ and $N_{{\nu}_s}-m_{{\nu}_s}$ planes. Notice that models with two massive  $0.5-1$~eV sterile neutrinos plus three sub-eV active states are perfectly allowed at $95\%$ CL by current Cosmic Microwave Background, galaxy clustering and Supernovae Ia data.
 Interestingly, these models are precisely the ones which, with the new reactor fluxes prediction, improve considerably the global fit to short baseline data. While we have not checked directly the results of Ref.~\cite{Kristiansen:2011mp} in which models with non zero curvature and two extra sterile neutrinos seem to exclude $w=-1$ at $95\%$ CL,
 it seems plausible that non standard $\Lambda$CDM cosmologies with sterile neutrino species provide also a very good fit to cosmological data. Big Bang Nucleosynthesis constraints could compromise the viability of these models if the two sterile neutrino states are fully thermal. If the BBN constraints are obtained using the helium fraction measurements $Y_p$ from \cite{aver} exclusively, two extra sterile neutrino states are perfectly allowed ($N_{{\nu}_s} < 2.3$ at 95\% CL). Even after combining these data with Deuterium measurements, two massive neutrino states are only marginally excluded  ($N_{{\nu}_s} < 1.7$ at 95\% CL). The tightest bound on the number of sterile neutrino species arises when helium measurements from Ref.~\cite{it} are combined with Deuterium data. Further developments in BBN determinations of light element abundances may have a large impact in further constraining the number of sterile neutrino species $N_{{\nu}_s}$.

\section{Acknowledgments}
RdP is supported by FP7-IDEAS-Phys.LSS 240117.
O.M. is supported by AYA2008-03531 and the Consolider Ingenio project CSD2007-00060.
We acknowledge support from the MICINN-INFN agreements ACI2009-1049, ACI-D-2011-0783 and 
INFN2008-016.
\appendix
\section{$\Lambda$CDM results}
\label{sec:lcdm}
Table \ref{tab:LCDM} summarizes the main results obtained in \cite{us} for a standard $\Lambda$CDM cosmology, for the shake of comparison with the results obtained within the other cosmological scenarios explored here.
\begin{table*}[ht]
\caption{1D marginalized 95\% CL bounds on $N_{\nu_s}$, $m_{\nu_s}$ and $m_\nu$  using the two combinations of data sets described in the text (r1 refers to ``run 1'' and r2 refers to ``run 2'', respectively) for a $\Lambda$CDM cosmology, see Ref.~\cite{us}. We also show the constraints after combining the results of ``run 2'' with those coming from different measurements of BBN light element abundances.}

\label{tab:LCDM}
\begin{center}\tabcolsep=2mm
\begin{tabular}{ c c c c c c c}
\hline\hline
 Parameter &  95\% CL(r1) &  95\% CL(r2) &  $Y^1_p$     &     $Y^2_p$     &     $Y^1_p+D$     & $Y^2_p+D$  \\
 \hline
$N_{\nu_s}$ & $<4.1$ & $<3.2$ & $<2.3$   &   $<1.7$   &   $<1.7$   &   $<1.4$ \\
 $m_{\nu}$\ [eV] & $<0.30$ & $<0.20$ & $<0.17$   &   $<0.15$   &   $<0.15$ &  $<0.15$\\
 $m_{\nu_s}$\ [eV] & $<0.46$ & $<0.50$ & $<0.62$   &   $<0.67$   &   $<0.69$ & $<0.68$\\
\hline\hline
\end{tabular}
\end{center}
\end{table*}




\begin{thebibliography}{99}

\frenchspacing

\bibitem{GonzalezGarcia:2007ib}
  M.~C.~Gonzalez-Garcia and M.~Maltoni,
  Phys.\ Rept.\  {\bf 460}, 1 (2008)
  [arXiv:0704.1800 [hep-ph]].
\bibitem{Lesgourgues:2006nd}
  J.~Lesgourgues and S.~Pastor,
  Phys.\ Rept.\  {\bf 429}, 307 (2006)
  [arXiv:astro-ph/0603494].
\bibitem{Komatsu:2010fb}
  E.~Komatsu {\it et al.}  [WMAP Collaboration],
  Astrophys.\ J.\ Suppl.\  {\bf 192}, 18 (2011)
  [arXiv:1001.4538 [astro-ph.CO]].
\bibitem{Reid:2009nq}
  B.~A.~Reid, L.~Verde, R.~Jimenez and O.~Mena,
  JCAP {\bf 1001}, 003 (2010)
  [arXiv:0910.0008 [astro-ph.CO]].
\bibitem{Hamann:2010pw}
  J.~Hamann, S.~Hannestad, J.~Lesgourgues, C.~Rampf and Y.~Y.~Y.~Wong,
  JCAP {\bf 1007}, 022 (2010)
  [arXiv:1003.3999 [astro-ph.CO]].
\bibitem{Mangano:2006ur}
  G.~Mangano, A.~Melchiorri, O.~Mena, G.~Miele and A.~Slosar,
  JCAP {\bf 0703}, 006 (2007)
  [arXiv:astro-ph/0612150].
\bibitem{Hamann:2007pi}
  J.~Hamann, S.~Hannestad, G.~G.~Raffelt and Y.~Y.~Y.~Wong,
  JCAP {\bf 0708}, 021 (2007)
  [arXiv:0705.0440 [astro-ph]].
\bibitem{Mangano:2010ei}
  G.~Mangano, G.~Miele, S.~Pastor, O.~Pisanti and S.~Sarikas,
  JCAP {\bf 1103} (2011) 035
  [arXiv:1011.0916 [astro-ph.CO]].
  \bibitem{darkr} 
  M.~Archidiacono, E.~Calabrese and A.~Melchiorri,
  arXiv:1109.2767 [astro-ph.CO].
  
 \bibitem{Aguilar:2001ty}
  A.~Aguilar {\it et al.}  [LSND Collaboration],
  Phys.\ Rev.\  D {\bf 64}, 112007 (2001)
  [arXiv:hep-ex/0104049].
\bibitem{Sorel:2003hf}
  M.~Sorel, J.~M.~Conrad and M.~Shaevitz,
  Phys.\ Rev.\  D {\bf 70}, 073004 (2004)
  [arXiv:hep-ph/0305255].
\bibitem{Karagiorgi:2006jf}
  G.~Karagiorgi, A.~Aguilar-Arevalo, J.~M.~Conrad, M.~H.~Shaevitz, 
  K.~Whisnant, M.~Sorel and V.~Barger,
  Phys.\ Rev.\  D {\bf 75}, 013011 (2007)
  [Erratum-ibid.\  D {\bf 80}, 099902 (2009)]
  [arXiv:hep-ph/0609177].
\bibitem{AguilarArevalo:2007it}
  A.~A.~Aguilar-Arevalo {\it et al.}  [The MiniBooNE Collaboration],
  Phys.\ Rev.\ Lett.\  {\bf 98}, 231801 (2007)
  [arXiv:0704.1500 [hep-ex]].
\bibitem{AguilarArevalo:2009xn}
  A.~A.~Aguilar-Arevalo {\it et al.}  [MiniBooNE Collaboration],
  Phys.\ Rev.\ Lett.\  {\bf 103}, 111801 (2009)
  [arXiv:0904.1958 [hep-ex]].
\bibitem{Karagiorgi:2009nb}
  G.~Karagiorgi, Z.~Djurcic, J.~M.~Conrad, M.~H.~Shaevitz and M.~Sorel,
  Phys.\ Rev.\  D {\bf 80}, 073001 (2009)
  [Erratum-ibid.\  D {\bf 81}, 039902 (2010)]
  [arXiv:0906.1997 [hep-ph]].

\bibitem{Mueller:2011nm}
  T.~.A.~Mueller, D.~Lhuillier, M.~Fallot, A.~Letourneau, S.~Cormon, M.~Fechner, L.~Giot, T.~Lasserre {\it et al.},
  Phys.\ Rev.\  {\bf C83}, 054615 (2011).
  [arXiv:1101.2663 [hep-ex]].
\bibitem{Mention:2011rk}
  G.~Mention, M.~Fechner, T.~Lasserre, T.~A.~Mueller, D.~Lhuillier, M.~Cribier and A.~Letourneau,
  arXiv:1101.2755 [hep-ex].
\bibitem{Kopp:2011qd}
  J.~Kopp, M.~Maltoni, T.~Schwetz,
  Phys.\ Rev.\ Lett.\  {\bf 107}, 091801 (2011).
  [arXiv:1103.4570 [hep-ph]].

\bibitem{Melchiorri:2008gq}
  A.~Melchiorri, O.~Mena, S.~Palomares-Ruiz, S.~Pascoli, A.~Slosar and M.~Sorel,
  JCAP {\bf 0901}, 036 (2009)
  [arXiv:0810.5133 [hep-ph]].
\bibitem{Acero:2008rh}
  M.~A.~Acero and J.~Lesgourgues,
  Phys.\ Rev.\  D {\bf 79} (2009) 045026
  [arXiv:0812.2249 [astro-ph]].
\bibitem{rt}
  J.~Hamann, S.~Hannestad, G.~G.~Raffelt, I.~Tamborra and Y.~Y.~Y.~Wong,
  Phys.\ Rev.\ Lett.\  {\bf 105}, 181301 (2010)
  [arXiv:1006.5276 [hep-ph]].

\bibitem{us}
  E.~Giusarma, M.~Corsi, M.~Archidiacono, R.~de Putter, A.~Melchiorri, O.~Mena, S.~Pandolfi,
  Phys.\ Rev.\  {\bf D83}, 115023 (2011).
  [arXiv:1102.4774 [astro-ph.CO]].
\bibitem{Hannestad:2005gj}
S.~Hannestad,
Phys.\ Rev.\ Lett.\  {\bf 95} (2005) 221301
[arXiv:astro-ph/0505551].
\bibitem{Hamann:2011ge}
  J.~Hamann, S.~Hannestad, G.~G.~Raffelt, Y.~Y.~Y.~Wong,
  JCAP {\bf 1109}, 034 (2011).
  [arXiv:1108.4136 [astro-ph.CO]].

\bibitem{Calabrese:2011hg} 
  E.~Calabrese, D.~Huterer, E.~V.~Linder, A.~Melchiorri and L.~Pagano,
  Phys.\ Rev.\ D {\bf 83}, 123504 (2011)
  [arXiv:1103.4132 [astro-ph.CO]].

\bibitem{LaVacca:2008mh}
  G.~La Vacca, S.~A.~Bonometto and L.~P.~L.~Colombo,
  New Astron.\  {\bf 14}, 435 (2009)
  [arXiv:0810.0127 [astro-ph]].
\bibitem{Gavela:2009cy}
  M.~B.~Gavela, D.~Hernandez, L.~L.~Honorez, O.~Mena and S.~Rigolin,
  JCAP {\bf 0907}, 034 (2009)
  [Erratum-ibid.\  {\bf 1005}, E01 (2010)]
  [arXiv:0901.1611 [astro-ph]].

\bibitem{Kristiansen:2011mp}
  J.~R.~Kristiansen, O.~Elgaroy,
  [arXiv:1104.0704 [astro-ph.CO]].
\bibitem{camb}
  A.~Lewis, A.~Challinor and A.~Lasenby,
  Astrophys.\ J.\  {\bf 538}, 473 (2000)
  [arXiv:astro-ph/9911177].
%
\bibitem{Lewis:2002ah}
  A.~Lewis and S.~Bridle,
  Phys.\ Rev.\  D {\bf 66}, 103511 (2002)
  [arXiv:astro-ph/0205436].
\bibitem{wmap7}
D.~Larson {\it et al.},
  Astrophys.\ J.\ Suppl.\  {\bf 192}, 16 (2011)
  [arXiv:1001.4635 [astro-ph.CO]].

\bibitem{beth}
  B.~A.~Reid {\it et al.},
  Mon.\ Not.\ Roy.\ Astron.\ Soc.\  {\bf 404}, 60 (2010)
  [arXiv:0907.1659 [astro-ph.CO]].

\bibitem{Riess:2009pu}
  A.~G.~Riess {\it et al.},
  Astrophys.\ J.\  {\bf 699}, 539 (2009)
  [arXiv:0905.0695 [astro-ph.CO]].
\bibitem{sn}
  R.~Amanullah {\it et al.},
  Astrophys.\ J.\  {\bf 716}, 712 (2010)
  [arXiv:1004.1711 [astro-ph.CO]].
  

\bibitem{aver}
 E.~Aver, K.~A.~Olive and E.~D.~Skillman,
  JCAP {\bf 1005}, 003 (2010)
  [arXiv:1001.5218 [astro-ph.CO]].
\bibitem{it}
  Y.~I.~Izotov and T.~X.~Thuan,
  Astrophys.\ J.\  {\bf 710}, L67 (2010)
  [arXiv:1001.4440 [astro-ph.CO]].
\bibitem{pettini}
  M.~Pettini, B.~J.~Zych, M.~T.~Murphy, A.~Lewis and C.~C.~Steidel,
  arXiv:0805.0594 [astro-ph].
\bibitem{iocco}
  O.~Pisanti, A.~Cirillo, S.~Esposito, F.~Iocco, G.~Mangano, G.~Miele and P.~D.~Serpico,
  Comput.\ Phys.\ Commun.\  {\bf 178}, 956 (2008)
  [arXiv:0705.0290 [astro-ph]].
\bibitem{hamann2}
  J.~Hamann, J.~Lesgourgues, G.~Mangano,
  JCAP {\bf 0803 } (2008)  004.
  [arXiv:0712.2826 [astro-ph]].
 \bibitem{Hou:2011ec} 
  Z.~Hou, R.~Keisler, L.~Knox, M.~Millea and C.~Reichardt,
  arXiv:1104.2333 [astro-ph.CO].
 \bibitem{arXiv:0803.0889} 
  K.~Ichikawa, T.~Sekiguchi and T.~Takahashi,
  Phys.\ Rev.\ D\ {\bf 78}, 083526  (2008)
  [arXiv:0803.0889 [astro-ph]].
  
   \bibitem{Chevallier:2000qy} 
  M.~Chevallier and D.~Polarski,
  Int.\ J.\ Mod.\ Phys.\ D\ {\bf 10}, 213  (2001)
  [gr-qc/0009008].
\bibitem{Linder:2002et} 
  E.~V.~Linder,
  Phys.\ Rev.\ Lett.\ \ {\bf 90}, 091301  (2003)
  [astro-ph/0208512].
  \bibitem{Albrecht:2006um} 
  A.~Albrecht, G.~Bernstein, R.~Cahn, W.~L.~Freedman, J.~Hewitt, W.~Hu, J.~Huth and M.~Kamionkowski {\it et al.},
  astro-ph/0609591.
\bibitem{Linder:2006sv} 
  E.~V.~Linder,
  Phys.\ Rev.\ D\ {\bf 73}, 063010  (2006)
  [astro-ph/0601052].
   \bibitem{Honorez:2010rr}
   L.~L.~Honorez, B.~A.~Reid, O.~Mena, L.~Verde and R.~Jimenez,
  JCAP\ {\bf 1009}, 029  (2010)
  [arXiv:1006.0877 [astro-ph.CO]].
\bibitem{Gavela:2010tm} 
  M.~B.~Gavela, L.~Lopez Honorez, O.~Mena and S.~Rigolin,
  JCAP {\bf 1011}, 044 (2010)
  [arXiv:1005.0295 [astro-ph.CO]].

\end{thebibliography}
\end{document}